# Laminar convective heat transfer of shear-thinning liquids in rectangular channels with longitudinal vortex generators

Amin Ebrahimi[1, †], Benyamin Naranjani[2], Shayan Milani[2], Farzad Dadras Javan[2]

1- Department of Materials Science & Engineering, Delft University of Technology, Mekelweg 2, 2628 CD Delft, The Netherlands.

2- Department of Mechanical Engineering, Faculty of Engineering, Ferdowsi University of Mashhad, Mashhad, P.O. Box 91775-1111, Iran.

† Corresponding author, Email: A.Ebrahimi@tudelft.nl, Phone: +31 15 2785682

**Abstract**

Heat and fluid flow in a rectangular channel heat sink equipped with longitudinal vortex generators have been numerically investigated in the range of Reynolds numbers between 25 and 200. Aqueous solutions of carboxymethyl cellulose (CMC) with different concentrations (200–2000 ppm), which are shear-thinning non-Newtonian liquids, have been utilised as working fluid. Three-dimensional simulations have been performed on a plain channel and a channel with five pairs of vortex generators. The channels have a hydraulic diameter of 8 mm and are heated by constant wall temperature. The vortex generators have been mounted at different angles of attack and locations inside the channel. The shear-thinning liquid flow in rectangular channels with longitudinal vortex generators are described and the mechanisms of heat transfer enhancement are discussed. The results demonstrate a heat transfer enhancement of 39–188% using CMC aqueous solutions in rectangular channels with LVGs with respect to a Newtonian liquid flow (*i.e.* water). Additionally, it is shown that equipping rectangular channels with LVGs results in an enhancement of 24–135% in heat transfer performance vis-à-vis plain channel. However, this heat transfer enhancement is associated with larger pressure losses. For the range of parameters studied in this paper, increasing the CMC concentration, the angle of attack of vortex generators and their lateral distances leads to an increase in heat transfer performance. Additionally, heat transfer performance of rectangular channels with longitudinal vortex generators enhances with increasing the Reynolds number in the laminar flow regime.

**1. Introduction**

Enhancing the thermal efficiency of heat exchangers is a challenging task to meet the heat removal capability needed for development of new devices with better performances. A number of designs and approaches have been proposed to passively enhance the heat transfer performance of cooling devices [1-9]. Equipping rectangular channels with vortex generators (VGs) has been demonstrated to be a promising method to passively augment the heat transfer performance [10-14]. Vortex generators with various shapes such as wing [15], winglet [16], rib [17, 18], pin fin [19] and surface protrusions [20-23] have been utilised for heat transfer enhancement applications. The pressure difference between two sides of VGs leads to flow separation from the side edges, which generates longitudinal, transverse and horseshoe vortices [24]. Formation of vortices intensifies fluid mixing and disrupts thermal boundary layer growth, which enhances the heat transfer performance [11, 25, 26]. Enhancing the heat removal capability of thermal systems is critical for developing new high-performance devices.

Apart from surface modifications, employing efficient coolants are expected to be a reasonable approach to enhance heat transfer performance of thermal systems. Liquid coolants are preferred over gaseous coolants for heat transfer enhancement applications because of their higher heat transfer coefficients [27, 28]. Non-Newtonian fluids are of high interest in practical applications since they can be made relatively easily compared with nanofluids [29]. It has been demonstrated that employing shear-thinning non-Newtonian fluids in thermal systems as working fluids can enhance the heat transfer performance of the system [30-34]. Aqueous solutions of carboxymethyl cellulose (CMC) are shear-thinning non-Newtonian liquids (also known as pseudoplastic liquids) that have been used for heat removal applications. Applications of non-Newtonian liquid coolants are usually bounded to the laminar flow regime because of low fluid velocity and high viscosity of the fluid [32, 35].

Velocity gradients and therefore shear stresses are larger in channels with LVGs in comparison with plain channels [14, 36]. It is known that shear-thinning behaviour influences the structure of vortices generated by VGs [29, 33, 37]. Therefore, the shear-thinning behaviour can result in a change in heat transfer performance of a heat exchanger [16, 26]. According to the best of the author's knowledge and to the reviewed literature, the influence of the non-Newtonian fluid flow behaviour on the thermo-hydraulic performance of rectangular channels equipped with LVGs is not addressed yet. More studies are essential to attain an insight into the effects of non-Newtonian fluid flow behaviour in channels equipped with LVGs and consequently on the heat transfer performance of channels. The primary aim of this study is to understand the influence of shear-thinning behaviour on fluid flow structure and heat transfer characteristics in rectangular channels equipped with LVGs. In the present study, three-dimensional numerical simulations have been performed to explore the shear-thinning power-law fluid flow structure and heat transfer in a rectangular channel with LVGs. To highlight the effect of shear-thinning behaviour on the heat transfer performance of a rectangular channel equipped with LVGs, CMC aqueous solutions with different CMC concentrations are scrutinised. Additionally, the influence of the shear rate on fluid flow structure and heat transfer performance is investigated by changing the angle of attack and the lateral distance of the LVGs. The non-Newtonian fluid flow structure and heat transfer characteristics are compared with a Newtonian fluid (*i.e.* water). The thermo-hydraulic performances of the channels with LVGs are also compared with a plain channel. The results presented in this paper may introduce new perspectives towards novel approaches to heat transfer enhancement in heat exchangers.

## 2. Model description

### 2.1. Computational domain

A rectangular channel with five pairs of LVGs was considered in the present study. A schematic diagram of the channel is shown in Figure 1. Six different configurations with VGs mounted at different angles of attack (*i.e.* α= 30°, 45° and 60°) and lateral distances (*i.e.* $d_t$= 5, 2.5 and 1.25 mm) were designed. The thickness of the VGs was idealised and supposed to be zero [38]. The geometrical parameters for different configurations are reported in Table 1. The heat and fluid flow were described in a three-dimensional Cartesian coordinate system, in which the mainstream was in the *z*-axis direction. The computational domain consisted of the inlet, main, and outlet zones. The inlet zone with adiabatic walls and the length of $L_{in}$ was defined to ensure the flow uniformity at the main zone entrance. The main zone encompasses five pairs of LVGs that were equally spaced in the mainstream direction. The solid walls were kept at constant temperature of 320 K in the main zone. The main zone was extended by adiabatic walls with the length of $L_{out}$ (*i.e.* outlet zone) to avoid any flow reversal at the outlet boundary.

**2.2. Physical model**

To highlight the differences between non-Newtonian and Newtonian coolants, aqueous solutions of carboxymethyl cellulose (CMC) were compared with a Newtonian liquid (*i.e.* water). Thermophysical properties of water and CMC aqueous solutions are presented in Table 2. Because of small temperature variations along the channel (less than 22 K), the material properties were assumed to be independent of temperature. To develop the mathematical model, the generation of longitudinal vortices was assumed to be quasi-steady [12] and the single-phase fluid flow to be laminar due to the low flow Reynolds number. Furthermore, it was assumed that the effects of body forces, radiation, and compressibility are negligible. Based on these assumptions, the continuum heat and fluid flow in the channel were modelled using the equations of mass, momentum, and energy conservation that are introduced as follows:

$$\nabla \cdot \vec{V} = 0, \qquad (1)$$

$$\rho\left(\vec{V} \cdot \nabla \vec{V}\right) = -\nabla p + \nabla \cdot \left[\mu\left(\nabla \vec{V} + \nabla^T \vec{V}\right)/2\right], \qquad (2)$$

$$\rho c_p (\vec{V} \cdot \nabla T) = k \nabla^2 T, \qquad (3)$$

where $\vec{V}$ is the fluid velocity vector, $\rho$ density [kg m$^{-3}$], $p$ the static pressure [Pa], $\mu$ dynamic viscosity [kg m$^{-1}$ s$^{-1}$], $c_p$ specific heat capacity [J kg$^{-1}$ K$^{-1}$], $T$ temperature [K], and k thermal conductivity [W m$^{-1}$ K$^{-1}$].

The power-law model was employed to describe the viscosity of shear-thinning fluids as a function of strain rate ($\dot{\gamma}$). The power-law model is expressed in Eq. 4, where $n$ and $K$ represent the power-law index [-] and the consistency index [kg m$^{-1}$ s$^{2-n}$], respectively. The strain rate and the second invariant of the rate-of-deformation tensor are interrelated [39]. The flow parameters of CMC aqueous solutions with different CMC concentrations are presented in Table 3.

$$\mu(\dot{\gamma}) = K\dot{\gamma}^{n-1} \tag{4}$$

The boundary conditions at the channel inlet, outlet, solid walls and symmetry plane are mathematically introduced as follows:

Inlet boundary: $u = v = 0$, $w = V_{in}$, $T = T_{in} = 298$ K,

Outlet boundary: $\dfrac{\partial u}{\partial z} = \dfrac{\partial v}{\partial z} = \dfrac{\partial w}{\partial z} = 0$, $\dfrac{\partial T}{\partial z} = 0$,

Symmetry plane: $\dfrac{\partial v}{\partial x} = \dfrac{\partial w}{\partial x} = 0$, $\dfrac{\partial T}{\partial x} = 0$, $u = 0$,

Heated walls (solid walls in the main zone): $u = v = w = 0$, $T = T_{wall} = 320$ K,

Adiabatic walls (solid walls in the inlet and the outlet zones): $u = v = w = 0$, $\dfrac{\partial T}{\partial y} = 0$,

**2.3. Numerical methodology**

Three-dimensional simulations were performed by solving the governing equations with the prescribed boundary conditions. Only half of the physical domain (*i.e.* the hatched region in Figure 1(b)) was considered in calculations because of the symmetric arrangement of the channel. The computational domain was discretized using structured non-uniform hexahedral grids. The computational grid is shown in Figure 2. Smaller grid sizes were used near the solid walls as well as the downstream of the VGs because of the higher flow gradients in those regions. The conservation equations of mass, momentum, and energy were discretized using the finite-volume approach. The SIMPLEC algorithm [40] was used to treat the pressure-velocity coupling. The convective terms and the diffusion terms of the governing equations were discretized with the upwind scheme both with second order accuracy. The convergence criteria for the continuity, momentum and the energy equations were defined to reach the scaled residues of $10^{-8}$, $10^{-8}$ and $10^{-10}$, respectively. The solver was built on top of an open-source flow solver, OpenFOAM [41]. All simulations were executed in parallel on three cores of an Intel Core i7-3520M processor (2.90 GHz).

**2.4. Data reduction**

The following dimensionless numbers are used to construct a framework of result presentation. The Reynolds number (Re) based on the hydraulic diameter of the channel ($D_h$) is defined as follows:

$$\text{Re} = \frac{\rho V_{in}^{2-n} D_h^n}{K}, \tag{5}$$

$$D_h = \frac{2WH}{W+H}, \tag{6}$$

It should be noted that *n* is equal to one for Newtonian fluids and therefore *K* represents the dynamic viscosity of the Newtonian liquid.

The Nusselt number (Nu), Prandtl number (Pr), required pumping power ($P_{pump}$), and Fanning friction factor (*f*) can be calculated using the following equations.

$$\text{Nu} = \frac{hD_h}{k}, \tag{7}$$

$$h = \frac{\dot{m}c_p(T_{out} - T_{in})}{A_h \Delta T}, \tag{8}$$

$$\Delta T = \frac{(T_{wall} - T_{in}) - (T_{wall} - T_{out})}{\ln[(T_{wall} - T_{in})/(T_{wall} - T_{out})]}, \tag{9}$$

$$\text{Pr} = \frac{c_p K (V_{in}/D_h)^{n-1}}{k}, \tag{10}$$

$$P_{pump} = Q \cdot \Delta p, \tag{11}$$

$$f = \frac{2\Delta p}{\rho V_{in}^2} \frac{D_h}{L_m}, \tag{12}$$

$$\Delta p = (\bar{p}_{out} - \bar{p}_{in}), \tag{13}$$

$$\bar{p} = \frac{\int p dA}{\int dA}, \tag{14}$$

where $\dot{m}$ is the mass flow rate of coolant [kg s$^{-1}$], $A_h$ the surface area of heated walls [m$^2$], $T_{out}$ the mass-weighted average of temperature at the outlet [K], and $Q$ the volumetric flow rate [m$^3$ s$^{-1}$]. The total pressure drop (Δ*p*) [Pa] was evaluated using $\bar{p}_{out}$ and $\bar{p}_{in}$, which are area-weighted static pressures at the exit and the entrance of the main zone, respectively.

The overall performance of the channels should be evaluated considering both the heat transfer and the friction loss. Therefore, the overall performance of the channels (η) is measured using a well-accepted performance evaluation parameter [14, 21, 32, 33, 42-44]. In Eq.15, Nu$_m$ is the mean Nusselt number and the subscript "s" stands for the plain channel. Cases with η>1.0 show better overall performances compared with the equivalent plain channel.

$$\eta = \frac{\text{Nu}_m}{\text{Nu}_{m,s}} \left( \frac{f_s}{f} \right)^{1/3} \tag{15}$$

## 3. Grid independence test and the solver verification

### 3.1. Grid independence test

Five meshes with a different number of cells were generated to investigate the sensitivity of the results to the cell size and to determine the minimum number of required cells to obtain reasonable results. The case of water flow in the channel with α=45° and Re=200 was considered for the grid independence test. The results of the grid independence test are compared with the results obtained from the largest grid (*i.e.* the mesh with 2.5×10$^6$ cells) and are presented in Table 4. Based on the results presented in Table 4, a mesh with 2×10$^6$ cells was selected for simulations considering both accuracy and computational costs.

### 3.2. Solver verification

The ability of the solver to reliably predict Newtonian fluid flow and heat transfer in rectangular channels with VGs was assessed in previous studies conducted by the authors [14, 16, 21]. The solver has been extended to model non-Newtonian fluid flow and heat transfer. Laminar convective heat transfer in a straight duct heated by constant wall temperature was considered to verify the solver. The duct has a length of 1.2 m and a square cross-section with the side length of 0.01 m. The fluid enters the duct with a constant temperature and velocity. The validity of the model was examined for both water and CMC aqueous solutions. In Table 5, the results of the present solver for the Reynolds number of 1000 were compared with the numerical results reported by Kurnia *et al.* [30]. Details assigned to the benchmark case can be found in [30]. The maximum deviation of the present numerical results from the reference data is 0.96 and 0.72% in the prediction of heat transfer rate and pressure drop, respectively, which demonstrates a reasonable agreement. This deviation from the referenced data can be attributed to differences in the grid size, numerical schemes and convergence criteria.

## 4. Results and discussion

Figure 3 shows the effects of the CMC concentration, the angle of attack and the lateral distance of the LVGs, and the Reynolds number on the mean Nusselt number ($Nu_m$). Compared with the plain channel, an augmentation of 24.17–134.83% in heat transfer performance is achieved by equipping the channels with LVGs. The intensified fluid mixing and secondary flow, and the disruption of thermal boundary layer growth result in an augmentation in heat transfer coefficient. It is seen that $Nu_m$ increases with increasing the Reynolds number for all the coolants considered in this study (Figure 3(a)). Increasing the Reynolds number causes a reduction in the thermal boundary layer thickness, which leads to higher heat transfer rates. Enlargement of the recirculation zone downstream of the VGs and strengthening the induced vortices at higher Reynolds numbers intensify the fluid mixing and the secondary flow in the channel [14, 36]. Increasing the Reynolds number results in higher velocity gradients and therefore leads to higher strain rates in the channel. For shear-thinning fluids, a higher strain rate leads to a reduction in viscosity. It makes the flow less stable, intensifies the fluid mixing and may initiate and/or augment the eddy generation and consequently enhances the heat transfer performance.

Figure 3(b) and Figure 3(c) show the influence of the angle of attack and the lateral distance of VGs on the mean Nusselt number, respectively. Based on the results shown in Figure 3(b), the larger the angle of attack of VGs, the higher the heat transfer performance of the channel. It is attributed to the strength of the secondary flow, the enlargement of the recirculation zone, and the flow destabilisation caused by the VGs with larger angles of attack. Increasing the angle of attack causes higher velocity gradients and strain rates in the channels. Higher strain rates result in a reduction in effective viscosity of shear-thinning fluids, which can lead to an augmentation of the fluid mixing in the channel and therefore the heat transfer performance of the channel. An increase in $Nu_m$ is observed with increasing $d_t$ (see Figure 3(c)). Decreasing the lateral distance of VGs leads to lower fluid velocities in the central region of the channel and higher velocities in the outer region. It causes a reduction in the size of the recirculation zone and the strength of the vortices and weakens the secondary flow. Additionally, due to the lower fluid velocities downstream of the VGs, the contribution of the convective heat transfer in the total heat transfer decreases by reducing the lateral distance of the VGs.

The results presented in Figure 3(d) indicate that, for the range of parameters studied in this paper, increasing the CMC concentration results in higher Nusselt numbers. The viscosity of coolants with higher CMC concentrations is more affected by the velocity gradients and therefore the heat and fluid flow are expected to be more influenced by the shear stresses generated by the VGs. The thermal conductivity of CMC aqueous solutions is higher than water, which enhances the heat absorption from the hot walls. CMC aqueous solutions with higher CMC concentrations have higher Prandtl numbers. Therefore, the momentum diffusivity is the dominant factor that governs the flow behaviour in comparison with the thermal diffusivity at higher CMC concentrations. It means that for the coolants with higher CMC concentrations, convection dominates the energy transportation in the channel compared with the conduction.

For the range of parameters studied here, an enhancement of 38.52–188.43% in the mean Nusselt number is recorded for the CMC aqueous solutions with respect to water. Figure 4 shows the pumping power required to drive the fluid flow through the channels. Employing CMC aqueous solutions as the coolant and equipping the channels with LVGs not only enhances the heat transfer performance but also increases the required pumping power. The total pressure drop in plain channels is mainly due to the friction of the walls, while for the channels with LVGs it depends on the friction of the walls, the form drag brought by the VGs, and the losses due to the secondary flow [14, 45, 46] (the contours of the dimensionless static pressure for different cases are shown in Figure S1 in the supplementary materials). The variations of the required pumping power with the Reynolds number is illustrated in Figure 4(a) for different CMC concentrations. Increasing the fluid velocity results in higher shear stresses and larger drag forces in the channel. Additionally, the interactions between the vortices and between the vortices and the solid walls lead to higher pressure losses at higher Reynolds numbers. On the basis of the results presented in Figure 4(b), increasing the angle of attack of the VGs leads to an increase in the required pumping power. It is attributed to the larger form drag due to the larger flow obstruction caused by the VGs with larger angles of attack and the changes in the flow pattern. Additionally, increasing the angle of attack of VGs destabilises fluid flow, increases fluid mixing and strengthen the generated vortices [11], which lead to an increase in the pressure penalty. Figure 4(c) shows the effect of the lateral distance of the VGs on the required pumping power. It is seen that the lateral distance of the VGs does not practically affect the required pumping power. It reveals that the form drag has the dominant contribution in the total pressure loss. It is observed that the bulk fluid viscosity is slightly higher in the channels that VGs were mounted closer to each other (*i.e.* smaller $d_t$), which increases the pressure drop. However, the weakened secondary flow reduces the pressure drop. The influence of CMC concentration on the required pumping power has been analysed and the results are shown in Figure 4(d). Higher pumping power is required to drive the flow of CMC solutions with larger CMC concentrations due to the higher bulk fluid viscosity. Based on Eq.4, the viscosity of CMC solutions is a function of strain rate and is more influenced by it at larger concentrations. Therefore, a change in CMC concentrations can change the hydraulic performance of the channel. Based on the results predicted by the numerical simulations, despite enhancing the heat transfer performance, more investment is needed to provide the pumping power required to drive the CMC aqueous solution flow with respect to the water flow.

Figure 5 shows the overall performance of the channels considered in this research as a function of Reynolds number. The results demonstrate an enhancement of 6.82–31.18% in overall performance vis-à-vis water for CMC aqueous solutions. The higher the CMC concentration, the better the overall performance of the heat sink. It is seen that the overall performance of all the cases with $d_t$=H enhances with increasing the Reynolds number. The overall performance enhances with decreasing the angle of attack of the VGs, which is attributed to the lower friction loss caused by smaller form drag. The results show that employing water as coolant inside the channel with α=60° is only efficient for Re>100. Additionally, increasing the lateral distance between the LVGs leads to higher overall performances. This can be further explained assessing the results shown in Figures 3 and 4.

The contours of temperature and the streamlines on a plane located at y/H=0.5 for different cases are presented in Figure 6 for the first three rows of the LVGs. The bulk fluid temperature decreases with increasing the Reynolds number due to the domination of the conductive by the convective heat transfer. The presence of the VGs disrupts the thermal boundary layer growth in the channel. Increasing the Reynolds number and/or the CMC concentration lead to a reduction in the thermal boundary layer thickness. The size of the recirculation zone decreases with decreasing the Reynolds number, the CMC concentration and the angle of attack of the LVGs. Higher fluid temperatures are found downstream of the LVGs due to the low fluid velocities in the recirculation zone. The fluid in the recirculation zone has time to absorb heat from the walls and thus the conductive heat transfer dominates the convection in the recirculation zone downstream of the VGs. Higher fluid temperatures are observed downstream of the VGs that are mounted closer to each other, which is due to the lower fluid velocity in those regions. The fluid velocity in the recirculation zone decreases with decreasing the lateral distance of the VGs because of the lower fluid velocity and the induced shear stress in the central region of the channel between the VGs. The temperature contours on a cross-section normal to the mainstream and located at the exit of the main zone (*i.e. z*/H =30) are shown in Figure 7 for different cases. Based on the secondary flow vectors, it is seen that increasing the angle of attack and the lateral distance of the VGs strengthens the secondary flow. The secondary flow induced by the generated vortices transfers the hot fluid adjacent to the channel walls towards the central region of the channel and the cold fluid in the central region to the hot walls. The strength of the generated vortices decreases moving towards the channel outlet. This churning fluid motion increases the temperature gradient and disrupts the thermal boundary layer growth resulting in higher heat transfer in the channel.

Figure 8 shows the contours of strain rate in the channels with LVGs. In this Figure, $\mu_m$ represents the volume-averaged fluid viscosity. It is seen that the presence of VGs causes higher velocity gradients in the fluid zone. The free shear layer generated by the LVGs significantly influences the flow pattern in the channel. Increasing the angle of attack of the VGs causes an increase in the shear rate and forms larger recirculation zones. Decreasing the Reynolds number reduces the velocity gradients in the fluid zone and hence decreases the magnitude of the strain rate and the size of the recirculation zone in the channel. The bulk fluid viscosity in the channel increases with increasing the CMC concentration. The higher degree of dependency of the fluid viscosity on the strain rate for the coolants with higher CMC concentrations results in local variations of the viscosity and changes the flow pattern in the channel and particularly downstream of the VGs. Consequently, the cores of the vortices become closer to the heated surface of the VGs leading to the heat transfer enhancement in the channel with CMC solutions of higher CMC concentrations. Decreasing the lateral distance of the VGs causes a reduction in the shear rate in the central region of the channel. It leads to lower fluid velocities in the recirculation zone downstream of the VGs and a weak secondary flow, which results in a lower heat transfer rates.

## 5. Conclusions

Three-dimensional simulations were conducted to investigate laminar convective heat transfer of shear-thinning liquids in rectangular channels with and without longitudinal vortex generators. Water and aqueous solutions of carboxymethyl cellulose (CMC) were selected as working fluid. The influences of the CMC concentrations, the angle of attack and the lateral distance of the vortex generators and the Reynolds number on heat and fluid flow pattern are studied. The results obtained from the present model show a reasonable agreement with available experimental and numerical data under the steady-state and laminar flow assumptions. Based on the results, the following conclusions are drawn.

Employing CMC aqueous solutions as the coolant in rectangular channels enhances the heat transfer performance. A further enhancement is achievable by equipping channels with longitudinal vortex generators. This heat transfer enhancement is associated with larger pressure losses, which makes these coolants suitable for applications that pumping power is not a snag. The heat transfer augmentation is mainly attributed to secondary flow, fluid mixing, and disruption of thermal boundary layer growth caused by the generated vortices. The strain rate, which is influenced by the shear layer generated by the vortex generators, can significantly affect the heat and fluid flow in the channel. It can be concluded that for the range of parameters studied in this paper, increasing the angle of attack and the lateral distance of the vortex generators leads to a larger strain rate in the fluid zone, stronger secondary flow and intensified fluid mixing and hence results in higher heat transfer rates.

Since the shear rates are more prominent in channels with LVGs, heat and fluid flow in channels with LVGs are more sensitive to the CMC concentrations compared with plain channels. Considering both the heat transfer and the friction loss, the overall performance of rectangular channels with LVGs is enhanced by using CMC aqueous solutions as the coolant. The ease and reliability of utilisation and maintenance are the main advantages of the proposed design. Since the shear rates are generally higher in channels with VGs, utilising shear-thinning fluids as the working medium provides an opportunity for heat transfer enhancement in channels with VGs.

**Nomenclature**

| | |
|---|---|
| $A_h$ | The surface area of heated walls, m$^2$ |
| $c_p$ | Specific heat capacity, J kg$^{-1}$ K$^{-1}$ |
| $D_h$ | Hydraulic diameter, m |
| $f$ | Fanning friction factor, - |
| $h$ | Convective heat transfer coefficient, W m$^{-2}$ K$^{-1}$ |
| k | Thermal conductivity, J m$^{-1}$ K$^{-1}$ |
| $K$ | Consistency index, kg m$^{-1}$ s$^{2-n}$ |
| $n$ | Power-law index, - |
| Nu | Nusselt number, - |
| $p$ | Static pressure, Pa |
| $P_{pump}$ | Pumping power, watt |
| $Q$ | Volumetric flow rate, m$^3$ s$^{-1}$ |
| Re | Reynolds number, - |
| $T$ | Temperature, K |
| $u$, $v$, $w$ | Velocity vector components |
| $x$, $y$, $z$ | Cartesian coordinates |
| $\alpha$ | Angle of attack, ° |
| $\eta$ | Overall performance, - |
| $\mu$ | Dynamic viscosity, Pa s |
| $\rho$ | Density, kg m$^{-3}$ |
| $\dot{m}$ | Mass flow rate, kg s$^{-1}$ |
| $\dot{\gamma}$ | Strain rate, s$^{-1}$ |

Acronyms

| | |
|---|---|
| CMC | carboxymethyl cellulose |
| VG | Vortex generator |
| LVG | Longitudinal vortex generator |

Subscripts

| | |
|---|---|
| in | Inlet |
| out | Outlet |
| wall | Wall |
| m | Mean |
| s | Plain channel |

**Supplementary materials**

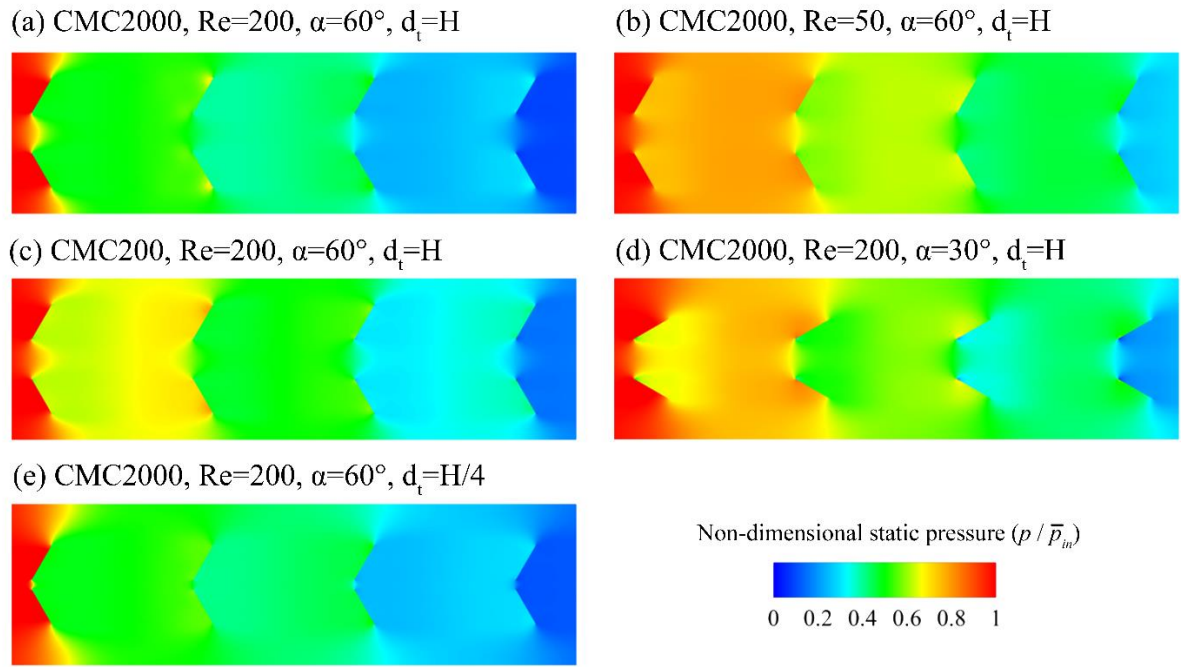

Figure S1. Contours of non-dimensional static pressure for different CMC concentrations, Reynolds numbers, angles of attack and lateral distances of VGs. (Contours are drawn in a plane located at y/H=0.5)

**References**

1. Hong, F. and P. Cheng, *Three dimensional numerical analyses and optimization of offset strip-fin microchannel heat sinks.* International Communications in Heat and Mass Transfer, 2009. **36**(7): p. 651-656.
2. Sui, Y., C.J. Teo, P.S. Lee, Y.T. Chew, and C. Shu, *Fluid flow and heat transfer in wavy microchannels.* International Journal of Heat and Mass Transfer, 2010. **53**(13–14): p. 2760-2772.
3. Bi, C., G.H. Tang, and W.Q. Tao, *Heat transfer enhancement in mini-channel heat sinks with dimples and cylindrical grooves.* Applied Thermal Engineering, 2013. **55**(1–2): p. 121-132.

**List of Tables**

Table 1. Geometrical parameters and characteristic dimensions of the channel and the vortex generators mounted in it.

Table 2. Thermophysical properties of water and CMC aqueous solutions (values taken from [32]).

Table 3. Flow parameters of CMC aqueous solutions for different CMC concentrations (values taken from [34]).

Table 4. The results of grid independence test for water flow in the channel with vortex generators positioned at the attack angle ($\alpha$) of 45° and Re=200.

Table 5. The results of the model verification for both Newtonian and non-Newtonian liquid flows in a duct with a square cross-section (Re=1000).

**List of Figures**

Figure 1. Schematic diagram of the channel and relevant geometrical parameters. (a) three-dimensional view of the channel and the vortex generators. (b) top-view of the channel and the vortex generators. (Only the hatched region was used for numerical simulations because of the symmetric flow pattern assumption)

Figure 2. The computational grid generated for the case of α=60° and $d_t$=H. Inlet and outlet zones are clipped for visualisation. To show the position of vortex generators clearly, their edges are shown with a finite thickness.

Figure 3. The effects of (a) Reynolds number, (b) angle of attack of the VGs, (c) lateral distance of VGs, and (d) CMC concentration on the mean Nusselt number. (Filled symbols: channels equipped with LVGs, Unfilled symbols: Plain channel)

Figure 4. The influence of (a) Reynolds number, (b) angle of attack of the VGs, (c) lateral distance of VGs, and (d) CMC concentration on the required pumping power. (Filled symbols: channels equipped with LVGs, Unfilled symbols: Plain channel)

Figure 5. The overall performance of the channels as a function of Reynolds number for different angles of attack of the VGs; (a) α=30°, (b) α=45°, (c) α=60°.

Figure 6. The influences of CMC concentration, Reynolds number, and the angle of attack and the lateral distance of the LVGs on the fluid flow and thermal fields. (Contours are shown in a plane located at *y*/H=0.5)

Figure 7. The contours of temperature and secondary flow vectors visualised on a cross-section located at the exit of the main zone (*i.e. z*/H=30). The CMC concentration, the Reynolds number, and the angle of attack and the lateral distance of the LVGs influence the thermal and the fluid flow fields.

Figure 8. Contours of strain rate for pseudoplastic liquid flow in the rectangular channel with LVGs. (Contours are shown in a plane located at $y/H=0.5$)

Table 1. Geometrical parameters and characteristic dimensions of the channel and the vortex generators mounted in it.

| Geometrical parameter | Value | Geometrical parameter | Value |
|---|---|---|---|
| $L_{in}$ | 10H | $d_{LVG}$ | 4H |
| $L_m$ | 20H | $d_t$ | H, H/2 and H/4 |
| $L_{out}$ | 10H | $L_{LVG}$ | H |
| W | 4H | α | 30°, 45° and 60° |
| $d_f$ | 1.5H | H | $5\times10^{-3}$ m |

Table 2. Thermophysical properties of water and CMC aqueous solutions (values taken from [32]).

| Water | | CMC aqueous solutions | |
|---|---|---|---|
| Property | Value | Property | Value |
| $\rho$ [kg m$^{-3}$] | 1000 | $\rho$ [kg m$^{-3}$] | 1000 |
| $c_p$ [J kg$^{-1}$ K$^{-1}$] | 4100 | $c_p$ [J kg$^{-1}$ K$^{-1}$] | 4100 |
| $k$ [J m$^{-1}$ K$^{-1}$] | 0.6076 | $k$ [J m$^{-1}$ K$^{-1}$] | 0.7 |
| $\mu$ [Pa s] | $9.55 \times 10^{-4}$ | $\mu$ [Pa s] | From Eq. 4 |

Table 3. Flow parameters of CMC aqueous solutions for different CMC concentrations (values taken from [34]).

| Concentration of CMC [ppm] | $n$ [-] | $K$ [kg m$^{-1}$ s$^{2-n}$] |
|---|---|---|
| 200 | 0.9270 | 0.00511 |
| 500 | 0.8229 | 0.00849 |
| 1000 | 0.7889 | 0.01235 |
| 2000 | 0.7051 | 0.02792 |

Table 4. The results of grid independence test for water flow in the channel with vortex generators positioned at the attack angle (α) of 45° and Re=200.

| Number of cells | Nu | % Diff. Nu | $f$ | % Diff. $f$ |
| --- | --- | --- | --- | --- |
| 450000 | 11.934 | 10.76 | 1.120 | 1.27 |
| 800000 | 11.644 | 8.06 | 1.120 | 1.27 |
| 1450000 | 11.394 | 5.74 | 1.119 | 1.18 |
| 2000000 | 10.829 | 0.50 | 1.108 | 0.18 |
| 2500000 | 10.775 | - | 1.106 | - |

Table 5. The results of the model verification for both Newtonian and non-Newtonian liquid flows in a duct with a square cross-section (Re=1000).

| Working fluid | Heat transfer rate [J s$^{-1}$] | | | Pressure drop in the duct [Pa] | | |
|---|---|---|---|---|---|---|
| | Kurnia *et al.* [30] | Present study | \|Difference\| [%] | Kurnia *et al.* [30] | Present study | \|Difference\| [%] |
| Water | 350.19 | 350.62 | 0.12 | 27.48 | 27.55 | 0.25 |
| CMC100 | 583.66 | 589.25 | 0.96 | 147.71 | 148.78 | 0.72 |
| CMC200 | 632.30 | 633.36 | 0.17 | 175.19 | 175.38 | 0.11 |
| CMC500 | 573.93 | 576.27 | 0.41 | 137.40 | 137.99 | 0.43 |
| CMC1000 | 622.57 | 622.82 | 0.04 | 168.32 | 168.77 | 0.27 |
| CMC2000 | 710.12 | 715.87 | 0.81 | 240.46 | 240.70 | 0.10 |

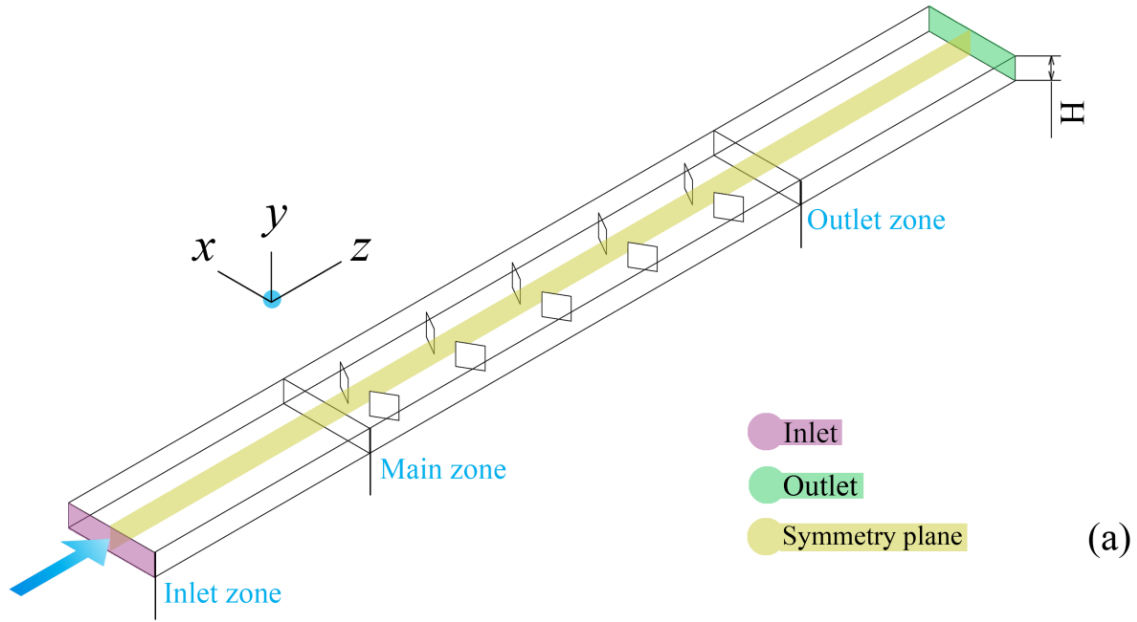

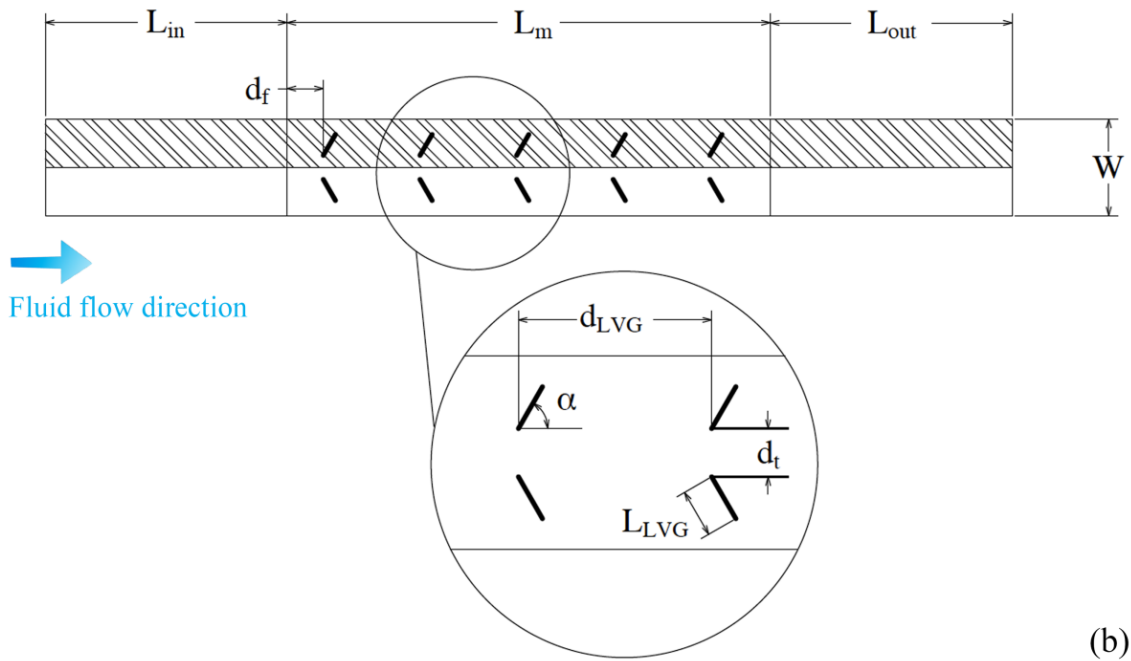

Figure 1. Schematic diagram of the channel and relevant geometrical parameters. (a) three-dimensional view of the channel and the vortex generators. (b) top-view of the channel and the vortex generators. (Only the hatched region was used for numerical simulations because of the symmetric flow pattern assumption)

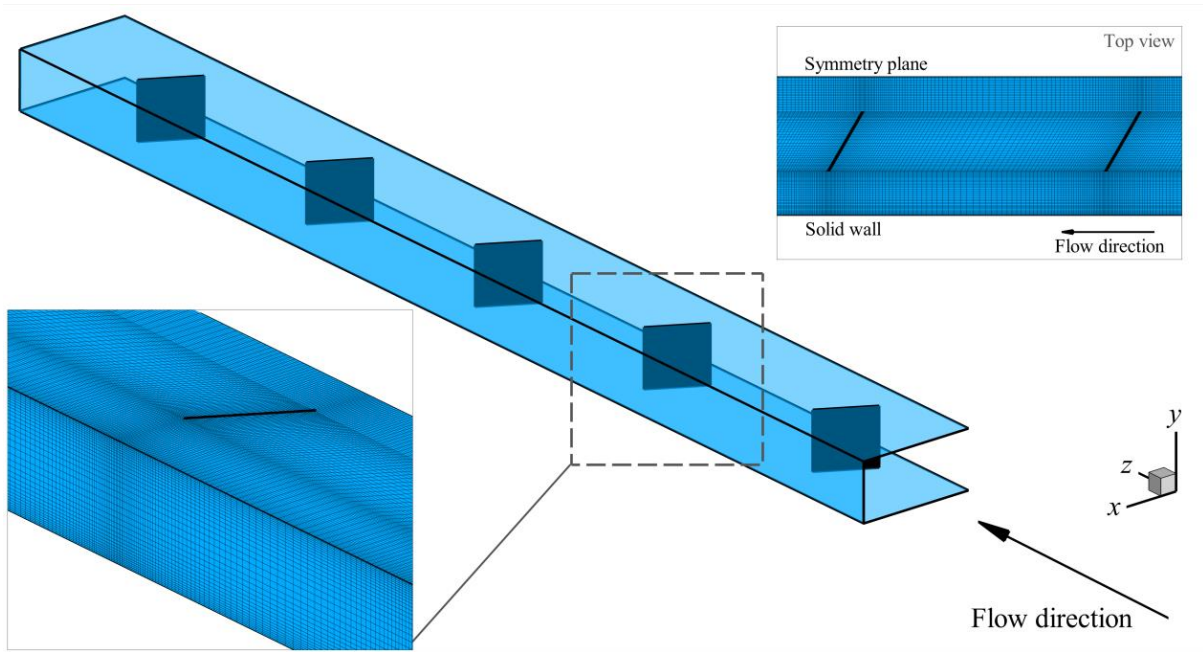

Figure 2. The computational grid generated for the case of $\alpha=60°$ and $d_t=H$. Inlet and outlet zones are clipped for visualisation. To show the position of vortex generators clearly, their edges are shown with a finite thickness.

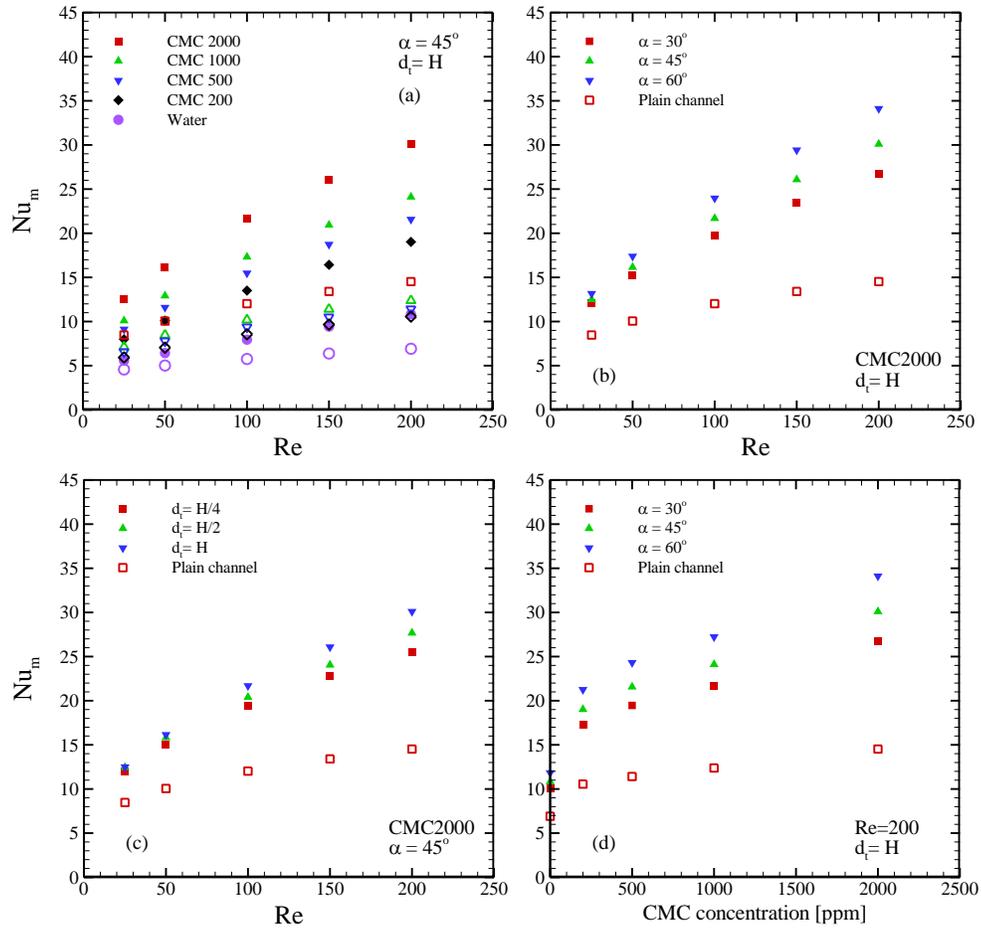

Figure 3. The effects of (a) Reynolds number, (b) angle of attack of the VGs, (c) lateral distance of VGs, and (d) CMC concentration on the mean Nusselt number. (Filled symbols: channels equipped with LVGs, Unfilled symbols: Plain channel)

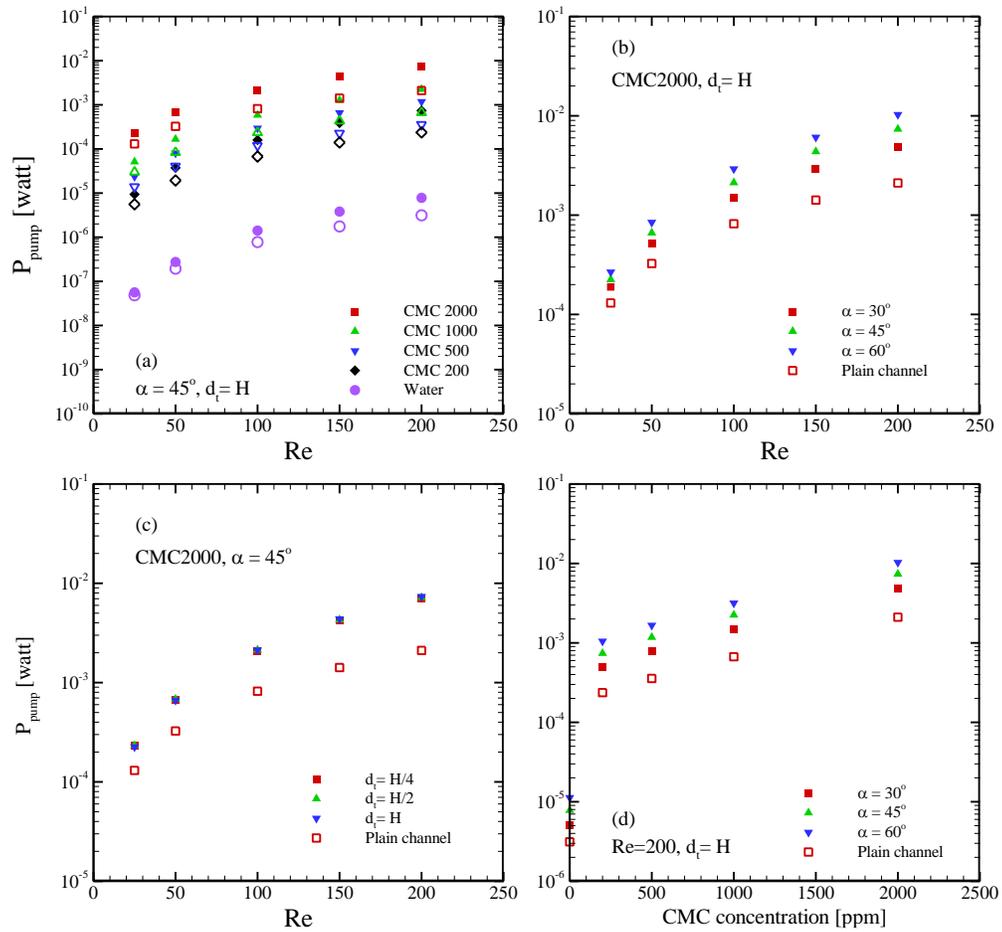

Figure 4. The influence of (a) Reynolds number, (b) angle of attack of the VGs, (c) lateral distance of VGs, and (d) CMC concentration on the required pumping power. (Filled symbols: channels equipped with LVGs, Unfilled symbols: Plain channel)

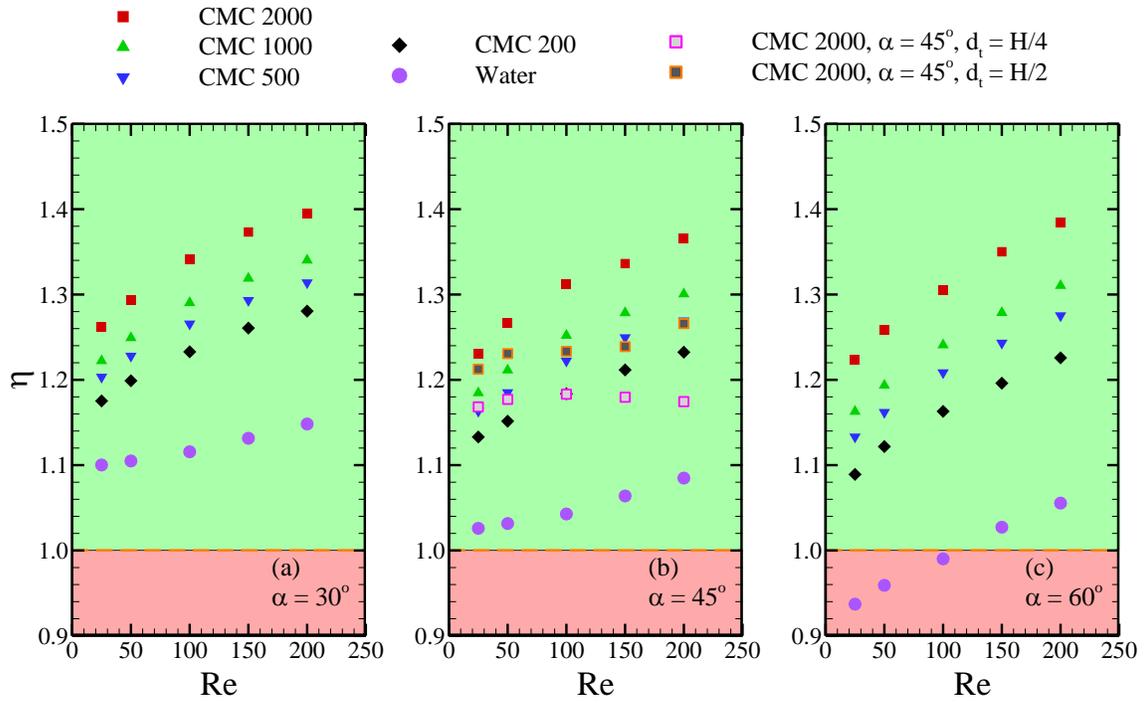

Figure 5. The overall performance of the channels as a function of Reynolds number for different angles of attack of the VGs; (a) α=30°, (b) α=45°, (c) α=60°.

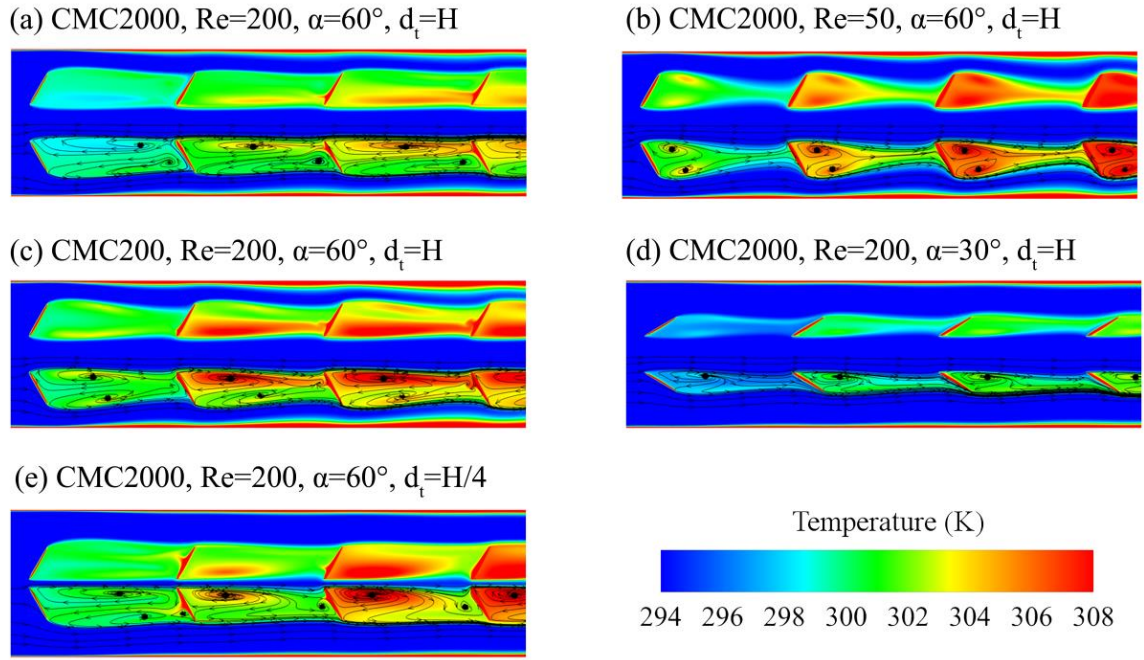

Figure 6. The influences of CMC concentration, Reynolds number, and the angle of attack and the lateral distance of the LVGs on the fluid flow and thermal fields. (Contours are shown in a plane located at $y/H=0.5$)

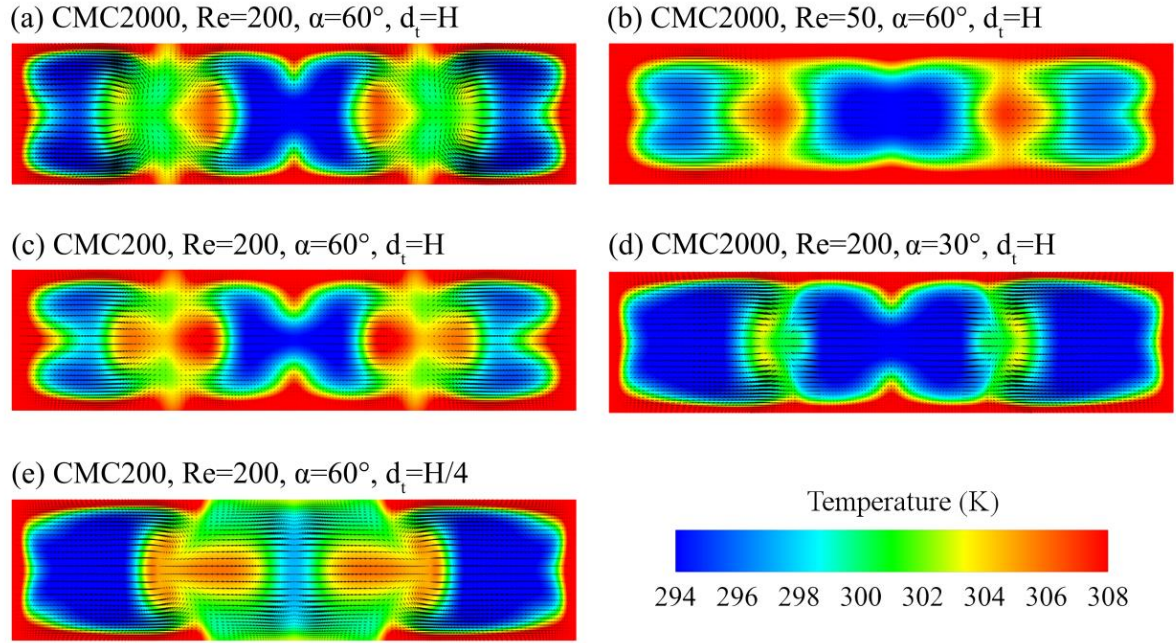

Figure 7. The contours of temperature and secondary flow vectors visualised on a cross-section located at the exit of the main zone (*i.e.* $z/H=30$). The CMC concentration, the Reynolds number, and the angle of attack and the lateral distance of the LVGs influence the thermal and the fluid flow fields.

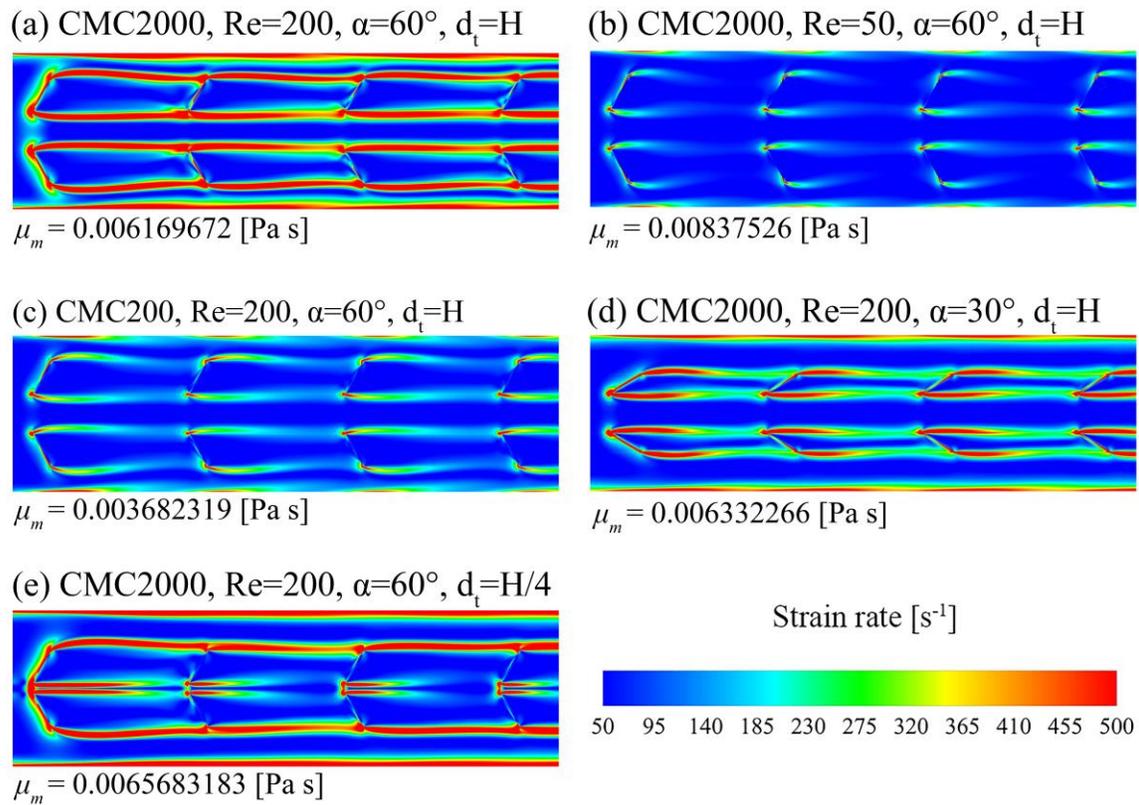

Figure 8. Contours of strain rate for pseudoplastic liquid flow in the rectangular channel with LVGs. (Contours are shown in a plane located at $y/H=0.5$)